\documentclass[11pt]{article}
\usepackage{amsmath}
\usepackage{amsfonts,amssymb}
\usepackage{simplewick}

\begin{document}

\interdisplaylinepenalty=5000

\renewcommand{\theequation}{\thesection.\arabic{equation}}
\let\ssection=\section
\renewcommand{\section}{\setcounter{equation}{0}\ssection}

\newcommand{\nl}{\nonumber\\}
\newcommand{\nnl}{\nl[6mm]}
\newcommand{\nle}{\nl[-2.5mm]\\[-2.5mm]}
\newcommand{\nlb}[1]{\nl[-2.0mm]\label{#1}\\[-2.0mm]}
\newcommand{\ab}{\allowbreak}

\newcommand{\e}{{\mathrm e}}
\newcommand{\mm}{{\mathbf m}}
\newcommand{\nn}{{\mathbf n}}
\newcommand{\rr}{{\mathbf r}}

\newcommand{\be}{\bes}
\newcommand{\ee}{\ees}
\newcommand{\bes}{\begin{eqnarray}}
\newcommand{\ees}{\end{eqnarray}}
\newcommand{\eens}{\nonumber\end{eqnarray}}

\renewcommand{\d}{\partial}

\newcommand{\bra}[1]{\langle{#1}|}
\newcommand{\ket}[1]{|{#1}\rangle}
\newcommand{\bracket}[2]{\langle{#1}{#2}\rangle}
\newcommand{\no}[1]{:{#1}:}

\newcommand{\eps}{\epsilon}
\newcommand{\si}{\sigma}
\newcommand{\ka}{\kappa}
\newcommand{\la}{\lambda}
\newcommand{\w}{\omega}

\newcommand{\dx}{d^dx\, }
\newcommand{\dy}{d^dy\, }

\newcommand{\vect}{{\mathfrak{vect}}}
\newcommand{\map}{{\mathfrak{map}}}
\newcommand{\gl}{{\mathfrak{gl}}}
\newcommand{\ssl}{{\mathfrak{sl}}}

\newcommand{\tr}{{\rm tr}}
\newcommand{\oj}{{\mathfrak g}}

\newcommand{\TT}{{\mathbb T}}
\newcommand{\RR}{{\mathbb R}}
\newcommand{\CC}{{\mathbb C}}
\newcommand{\ZZ}{{\mathbb Z}}
\newcommand{\NN}{{\mathbb N}}

\newcommand{\xmu}{\xi^\mu}
\newcommand{\ynu}{\eta^\nu}
\newcommand{\dmu}{\d_\mu}
\newcommand{\dnu}{\d_\nu}
\newcommand{\drho}{\d_\rho}
\newcommand{\dsi}{\d_\si}

\renewcommand{\L}{{\mathcal L}}
\newcommand{\J}{{\mathcal J}}
\newcommand{\T}{{\mathcal T}}
\newcommand{\Z}{{\mathcal Z}}

\newcommand{\summp}{\sum_{|\mm|\leq p}}
\newcommand{\sumnp}{\sum_{|\nn|\leq p}}

\newcommand{\mfrac}{\frac{(-)^{|\mm|}}{\mm!}}
\newcommand{\nfrac}{\frac{(-)^{|\nn|}}{\nn!}}
\newcommand{\xeta}{{\xi,\eta}}
\newcommand{\xX}{{\xi,X}}

\title{{New Derivation of Off-Shell Representations of the
 Multi-dimensional Affine and Virasoro Algebras}}

\author{T. A. Larsson \\
Vanadisv\"agen 29, S-113 23 Stockholm, Sweden\\
email: thomas.larsson@hdd.se}

\maketitle
\begin{abstract}
Algebras of currents and diffeomorphisms in arbitrary dimension
have extensions which generalize the affine and Virasoro algebras
on the circle. A large class of off-shell representations was discovered
in Comm. Math. Phys. {\bf 214} (2000) 469--491.
That paper is not so accessible due to a slightly non-standard normal 
ordering formalism and cumbersome $p$-jet calculations. 
The purpose of the present paper is to simplify the derivation 
using standard OPE methods and a more field-like formalism.
\end{abstract}

\vskip 1cm

\section{Introduction}

The algebra of diffeomorphisms on the circle has a non-trivial central
extension, the Virasoro algebra. A natural question is whether this
extension can be generalized to diffeomorphisms on higher-dimensional
manifolds, in particular on the $d$-dimensional torus. The answer is affirmative
and the multi-dimensional Virasoro algebra $Vir(d)$ was discovered 
during the 1990s \cite{Lar91,RM94}, together with many representations 
\cite{BB98,Bil97,Lar98,RM94}.

$Vir(d)$ for $d>1$ differs in many respects from the ordinary Virasoro algebra
$Vir(1)$. E.g., the Virasoro-like extension is not central unless $d=1$. In
$d$ dimensions, it is an extension by the module of closed $(d-1)$-forms.
When $d=1$, a closed zero-form is a constant function, and the extension is
central. In higher dimensions, the Lie algebra extension is still well-defined, 
but it does not commute with diffeomorphisms.

The off-shell representations are also essentially different. Fock
representations of $Vir(1)$ are obtained by quantizing fields on the circle.
However, this procedure does not generalize to higher dimensions, because
normal ordering gives rise to non-removable infinities. In some sense, this
reflects the fact that Quantum Field Theory is incompatible with 
general-covariant theories like gravity. To construct off-shell representations
of $Vir(d)$ with $d>1$, we must replace the fields with histories in the 
corresponding spaces of $p$-jets prior to quantization. After this step is
taken, we have a classical representation acting on finitely many functions
of a single variable, which can be quantized without the appearance of infinites. 

Unfortunately, working with $p$-jets is quite cumbersome. The purpose of the 
present paper is to simplify the calculations and to emphasize the close relation
to field theory. 
Locally, a $p$-jet is simply a Taylor expansion truncated at order $p$; 
in one dimension, the $p$-jet corresponding the field $\phi(x)$ is
\be
\phi(x)\rvert_p = \sum_{m=0}^p \frac1{m!} \phi_m (x-q)^m,
\label{pjetA}
\ee
where the Taylor coefficients $\phi_m = d^m \phi/dx^m(q)$. 

A Taylor series does not only depend on the function being
expanded, but also on the choice of expansion point $q$, to be
identified with the {\em observer's position}. This means that
(\ref{pjetA}) does not commute with the observer's momentum, i.e.
the canonical conjugate of $q$. 
To obtain an orthogonal set of canonical variables, we replace the
{\em absolute field} $\phi(x) \equiv \phi_A(x)$ with the corresponding
{\em relative field} $\phi_R(x) = \phi_A(x+q)$,
\be
\phi_R(x)\rvert_p = \sum_{m=0}^p \frac1{m!} \phi_m x^m,
\label{pjetR}
\ee
which clearly commutes with the observer's momentum.
To rephrase any expression in terms of relative fields, we substitute
$\phi_A \to \phi_R$ and $x \to x+q$. This step explicitly brings out 
observer-dependence. In particular, the multi-dimensional
Virasoro cocycles are non-central because they are functionals of 
the observer's trajectory $q(t)$, which does not commute with diffeomorphisms.

The results in \cite{Lar98} were formulated directly in terms of the
$p$-jet data, i.e. the Taylor coefficients $\phi_m$, the observer's
position $q$, and their canonical momenta. 
To facilitate calculations and make 
formulas resemble the corresponding field equations, in the
present paper we work with the full MacLaurin series (\ref{pjetR}) 
rather than the individual Taylor coefficients.
However, the underlying $p$-jet structure is still present, and manifests
itself in some places. We need to replace the $d$-dimensional delta 
function $\delta(x-y)$ with the corresponding $p$-jet delta function
$\delta_p(x,y)$. Unlike the ordinary delta function, this object can
be multiplied with itself in a meaningful way. That the square
of the ordinary delta function is infinite is one reason why the
multi-dimensional Virasoro algebra can not be obtained by quantizing
the original fields.

The passage to $p$-jets can be viewed from a slightly different 
perspective. Introduce a nilpotent number $\eps$ satisfying $\eps^p=0$,
replace all coordinates $x^\mu$ with $\eps x^\mu$, and set 
$\eps^m = 1$ for all $m\leq p$ at the end. 
This procedure automatically picks out the $p$-jet part, since
$\phi_R(\eps x) = \phi_R(\eps x)\rvert_p$.
Moreover, the product of $p$-jets is handled correctly, because
\be
\phi_R(\eps x)\rvert_p\, \psi_R(\eps y)\rvert_p
= \Big(\phi_R(\eps x)\psi_R(\eps y)\Big) \Big\rvert_p.
\ee
The use of nilpotent numbers is not necessary for evaluating bilinear products
of operators, because the canonical momentum will pick out 
the correct $p$-jet part of products of fields anyway, but the method
may be useful in other contexts.

A separate issue with \cite{Lar98} is that normal ordering was 
carried out using a slightly non-standard formalism. In the present
paper we use the standard operator product expansion (OPE) instead,
following the notation of \cite{FMS94}. 
This method is much simpler and should hopefully make
the results more accessible.

Before addressing the full diffeomorphism algebra, we start with the
simpler case of gauge transformations in Yang-Mills type theories, i.e.
the current algebra in $d$ dimensions. It admits a non-trivial extension,
which we call the multi-dimensional affine algebra $Aff(d,\oj)$. The same
algebra is called the {\em central extension} in \cite{EF94,PS88}, but we
avoid this name because the extension does not commute with
diffeomorphisms.

This paper is organized as follows. In the next section some necessary
formalism is established.
$Aff(d,\oj)$ and its off-shell representations are discussed in section 3.
The more complicated case of $Vir(d)$ is dealt with in the section
thereafter. In section 5 we note that the representations also admit an
intertwining action of an additional Virasoro algebra, which can be
identified as the algebra of reparameterizations of the observer's
trajectory. We can thus enlarge the symmetry to include reparametrizations
``for free'', i.e. without enlarging the modules.
The main results are summarized in section 6. The final section contains 
a brief discussion of the relevance of these algebras to physics. 
Heavy calculations are relegated to the appendices.

\section{ $p$-jet preliminaries }
\label{sec:pjet}

To do calculations with $p$-jets it is useful to introduce multi-indices.
Let $\mm = (m_0, \ab m_1, \ab ..., \ab m_{d-1})$, all $m_\mu\geq0$, be a
multi-index of length $|\mm| = \sum_{\mu=0}^{d-1} m_\mu$.
The factorial is $\mm! = m_0!m_1!...m_{d-1}!$, and the power is 
\be
x^\mm = (x^0)^{m_0} (x^1)^{m_1} ... (x^{d-1})^{m_{d-1}}.
\label{power}
\ee
Multi-indices can be added and subtracted,
\be
\mm+\nn = (m_0+n_0, m_1+n_1, ..., m_{d-1}+n_{d-1}).
\ee
The binomial coefficients are defined in the obvious way, i.e.
\break $\binom{\mm}{\nn} = \mm!/\nn!((\mm-\nn)!$.
Finally, let $\mu$ also denote the unit multi-index in the $\mu$:th direction,
i.e. $\mu_\nu = \delta_{\mu,\nu}$, and hence
$(\mm+\mu)! = m_\mu \mm!$.

Let $\phi_A(x)$ is an {\em absolute field}, where the coordinates $x = (x^\mu)$ 
are defined relative some fixed but arbitrary global origin. 
A $p$-jet is the Taylor expansion of a field $\phi_A(x)$ around the
observer's position $q^\mu$, truncated at order $p$:
\be
\phi_A(x)\rvert_p = \summp \frac1{\mm!} \phi_\mm (x-q)^\mm.
\label{phi_A}
\ee
The space of $p$-jets is spanned by the Taylor coefficients
$\phi_\mm$, $|\mm| \leq p$, and the expansion point $q^\mu$. 

Introduce the corresponding canonical momenta $\pi^\mm$ and $p_\mu$, 
satisfying the canonical commutation relations (CCR):
\be
[\phi_\mm, \pi^\nn] = i \delta^\nn_\mm, \qquad
[\phi_\mm, \phi_\nn] = [\pi^\mm, \pi^\nn] = 0,
\label{H-phi-pi}
\ee
and
\be
[q^\mu, p_\nu] = i\delta^\mu_\nu, \qquad
[q^\mu, q^\nu] = [p_\mu, p_\nu] = 0.
\label{H-q-p}
\ee
The observer's momentum $p_\mu$ does not commute
with absolute fields, because
\bes
[p_\mu, \phi_A(x)] &=& \summp \frac{i}{(\mm-\mu)!} \phi_\mm (x-q)^{\mm-\mu} 
\nl
&=& i\dmu\phi_A(x).
\ees
To obtain an independent set of canonical variables, we introduce the
corresponding {\em relative field} $\phi_R(x)$, where the coordinates $x$ are 
measured relative to the observer's position $q^\mu$, rather than relative to some
fixed global origin. Define
\be
\phi_R(x) \equiv \phi_A(x+q),
\ee
and hence
\be
\phi_A(x) = \phi_R(x-q).
\ee
The Taylor expansion of the relative field takes the form of a MacLaurin series:
\be
\phi_R(x) = \phi_R(x)\rvert_p = \summp \frac1{\mm!} \phi_\mm x^\mm.
\label{phiR}
\ee
For any field $A(x)$, we use the notation $A(x)\rvert_p$ to denote the 
corresponding $p$-jet.

Whereas (\ref{phiR}) can be viewed as a generating function for the Taylor
coefficient $\phi_\mm$, the natural generating function for the jet momenta
is a field in momentum space:
\be
\hat\pi_R(k) = \summp i^{|\mm|} \pi^\mm k^\mm.
\label{hatpi}
\ee
For infinite jets, the CCR (\ref{H-phi-pi}) becomes
\be
[\phi_R(x), \hat\pi_R(k)] = i\exp(ik\cdot x), \qquad (p = \infty).
\ee
Making an inverse Fourier transform of (\ref{hatpi}) we obtain 
the following expression for the canonical momentum in position space:
\be
\pi_R(x) = \summp (-)^{|\mm|} \pi^\mm \d_\mm\delta(x)
\label{piR}
\ee
The CCR in position space are thus
\be
[\phi_R(x), \pi_R(y)] = i\delta_p(x,y),
\ee
where
\be
\delta_p(x,y) = \summp \mfrac\, x^m\, \d_m\delta(y).
\label{delta_def}
\ee
is the {\em $p$-jet delta function}.
This is not symmetric,
\be
\delta_p(y,x) \neq \delta_p(x,y),
\label{delta_asym}
\ee
and it has the following smearing properties, which are proven in 
Appendix \ref{app:pjet}:
\bes
\int \dy f(y) \delta_p(x,y) &=& f(x) \rvert_p, 
\label{delta_smear}\\
\int \dy f(y) \d^x_\mu \delta_p(x,y) &=& \dmu f(x) \rvert_p, 
\label{partint_x}\\
\int \dy f(y) \d^y_\mu \delta_p(x,y) &=& -\dmu f(x) \rvert_p.
\label{partint_y}
\ees
for every smearing function $f(x)$. 

A crucial difference between the $p$-jet delta function and the ordinary field
delta function is that the product of the former with itself makes sense.
In appendix \ref{app:delta_product} we prove the following crucial results:
\bes
\delta_p(x,y)\, \delta_p(y,x) 
&=& \binom{d+p}{d} \delta(x)\delta(y), \nl
\d^x_\mu \delta_p(x,y)\, \delta_p(y,x) 
&\approx& -\binom{d+p}{d+1} \dmu \delta(x)\delta(y), 
\nle
\d^x_\mu \delta_p(x,y)\, \d^y_\nu \delta_p(y,x) 
&\approx& \binom{d+p+1}{d+2} \dnu\delta(x) \dmu\delta(y) \nl
&+& \binom{d+p}{d+2} \dmu\delta(x) \dnu\delta(y).
\eens
The first relation is a strict equality, but the two latter must be understood
in a weaker sense. They become equalities if the RHS is smeared with a function
$f(x)$ and the LHS is smeared with the function $f_0(x) = f(x) - f(0)$; in the
last relation an analogous subtraction is also necessary for the smearing function
$g(y)$. Fortunately, it is exactly this modified smearing that is necessary for
our purposes.

From the definition (\ref{piR}) and an integration by parts it follows that
the canonical momentum picks out the $p$-jet part of any field:
\be
\int \dx \pi_R(x)A(x) = \int \dx \pi_R(x)A(x)\rvert_p.
\label{piprop}
\ee
Another useful property of $p$-jets is that
\be
(A(x)\rvert_p B(x)\rvert_p)\rvert_p = (A(x)B(x))\rvert_p.
\label{pjetprop}
\ee
To prove (\ref{pjetprop}), we observe that both sides equal the double sum
\be
\sum_\mm \sum_\nn \frac1{\mm!\nn!} A_\mm B_\nn x^{\mm+\nn},
\ee
but the summation ranges seem different. In the LHS, the sum runs over $\mm$ and
$\nn$ which satisfy the joint condition $|\mm|\leq p$, $|\nn|\leq p$, and 
$|\mm+\nn|\leq p$, whereas in the RHS the only condition is $|\mm+\nn|\leq p$.
However, the two first conditions in the LHS are in fact redundant, because if
$|\mm+\nn|\leq p$, then $|\mm|$ and $|\nn|$ are both automatically $\leq p$. Hence the 
two sides of (\ref{pjetprop}) are indeed equal.

\section{ Multi-dimensional affine algebra }
\label{sec:Aff}

\subsection{ Classical representations }

Let $\oj$ be a Lie algebra with basis $J^a$, 
totally anti-symmetric structure constants $f^{abc}$, 
and Killing metric $\delta^{ab}$. Due to the existence of the Killing metric,
there is no need to distinguish between upper and lower $\oj$ indices.
The $\oj$ brackets are
\be
[J^a, J^b] = if^{abc} J^c.
\ee
Let the matrices $M^a$ form a basis for a finite-dimensional $\oj$ representation 
$M$, and let $y_M$ be the value of the second Casimir operator in this
representation, defined by
\be
\tr\,M^a M^b = y_M \delta^{ab}.
\label{Casimir}
\ee

The algebra of maps from the $d$-dimensional base manifold into $\oj$, 
$\map(d,\oj)$, is defined by the brackets
\be
{[}\J_X, \J_Y] &=& \J_{[X,Y]},
\ee
where $X = X^a(x) J^a$ is a $\oj$-valued function and 
$[X,Y] = if^{abc} X^a Y^b J^c$.

For each finite-dimensional $\oj$ representation $M$, 
the corresponding $\map(d,\oj)$ representation acts on $M$-valued fields:
\be
[\J_X, \phi(x)] = - X^a(x) M^a \phi(x) \equiv -X(x) \phi(x).
\ee
Let $\pi(x)$ be the canonical conjugate of $\phi(x)$, defined by the 
Heisenberg algebra
\bes
[\phi(x), \pi(y)] &=& i\delta(x-y), 
\nle
[\phi(x), \phi(y)] &=& [\pi(x), \pi(y)] = 0.
\eens
$\pi(x)$ transforms in the dual representation:
\be
[\J_X, \pi(x)] = X^a(x) \pi(x) M^a = \pi(x) X(x).
\ee
Hence the current algebra is generated by the operators
\be
\J_X = -i \int \dx \pi(x) X(x) \phi(x).
\label{JXclass}
\ee

\subsection{ Quantization in one dimension }

Let us first review how to build representations of the
ordinary affine algebra $Aff(1)$, using the OPE formalism \cite{FMS94}.
Let $\phi(z)$ be an $M$-valued field and let $\pi(z)$ be its canonical
momentum. The OPEs of the fundamental fields read
\bes
\phi(z)\pi(w) &\sim& \frac{1}{z-w}, 
\nlb{ope}
\phi(z)\phi(w) &\sim& \pi(z)\pi(w) \sim 0.
\eens
We denote the $z$ derivative by a dot, $\dot\phi(z) \equiv d\phi(z)/dz$,
to distinguish it from derivatives w.r.t. spacetime coordinates below.
Note that we have not identified momenta and velocities, $\pi(z) \neq \dot\phi(z)$.
In string theory fields obeying the definition (\ref{ope}) are often 
called ghosts, cf \cite{FMS94}, eqn. (6.238).

Equation (\ref{ope}) holds both for bosonic and fermionic fields, but the OPE of
the fields in opposite order depends on Grassman parity:
\bes
\pi(z)\phi(w) \sim -\frac{1}{z-w}, &\qquad& \hbox{bosonic}
\nle
\pi(z)\phi(w) \sim \frac{1}{z-w}, &\qquad& \hbox{fermionic}.
\eens
To treat both bosons and fermions at the same time, we summarize the OPE of the 
fundamental fields as
\be
\phi(z)\pi(w) \sim \mp \pi(z)\phi(w) \sim \frac{1}{z-w}.
\ee
Here and henceforth the upper sign refers to bosons and the lower sign to fermions.

The generators of $Aff(1,\oj)$ are
\be
J^a(z) = \no{ \pi(z) M^a \phi(z) },
\ee
where double dots denotes normal ordering.
The field transforms in $M$ and the canonical momentum
transforms in the dual representation:
\bes
J^a(z) \phi(w) &\sim& -\frac{M^a \phi(w)}{z-w}, \nle
J^a(z) \pi(w) &\sim& \frac{\pi(w)M^a}{z-w},
\eens
The affine algebra 
\be
J^a(z) J^b(w) &\sim& if^{abc} \frac{J^c(w)}{z-w} + \frac{k\delta^{ab}}{(z-w)^2},
\ee
where the central charge $k = \mp y_M$ is given by the value of the
second Casimir in (\ref{Casimir}).
In terms of the smeared generators 
\be
\J_X = \frac1{2\pi i} \oint dz\, X^a(z) J^a(z),
\ee
the affine algebra takes the form
\bes
[\J_X, \J_Y] &=& \oint_0 dw\, \oint_w dz\, J^a(z) J^b(w) \nle
&=& \J_{[X,Y]} + \frac{k}{2\pi i} \oint dw\, \dot X^a(w) Y(w).
\eens

\subsection{ Na\"ive field quantization }

Let us now attempt to repeat the construction in the previous subsection 
for $Aff(d,\oj)$. 
In several dimensions there is no preferred time direction that 
defines the normal order. One could define
a lexicographical order as is done for finite-dimensional Lie algebras,
but this is problematic for several reasons, e.g. because there is
no guarantee that different orderings will lead to equivalent representations;
the Stone-von Neumann theorem does not hold in infinite dimensions.

Instead, we explicitly introduce an extra coordinate $z$ to 
define normal ordering. One way to think about this is that we have fixed a foliation, 
$t$ is the time coordinate, and the complex coordinate $z = \exp(it)$;
$x$ denotes the spatial coordinates.
$\map(d,\oj)$ is then the algebra of spatial gauge transformations. 
Another interpretation is that $t$ is a time-like parameter along the 
observer's trajectory, which should be identified with the time coordinate
$x^0$ at a later stage. 

Either way, the fundamental fields now depend on an extra complex variable, 
and the OPEs become
\bes
\phi(x,z)\pi(y,w) &\sim& \mp \pi(x,z)\phi(y,w)\, \sim\, \frac{1}{z-w}\, \delta(x-y), 
\nlb{ope2}
\phi(x,z)\phi(y,w) &\sim& \pi(x,z)\pi(y,w)\, \sim\, 0.
\eens
The smeared generators,
\be
J_X(z) = \int \dx \no{ \pi(x,z) X(x,z) \phi(x,z) },
\label{JXzabs}
\ee
act as follows on the fundamental fields
\bes
J_X(z) \phi(x,w) &\sim& \frac{-1}{z-w} X(x,w)\phi(x,w), \nle
J_X(z) \pi(x,w) &\sim& \frac1{z-w} \pi(x,w)X(x,w).
\eens

We now encounter a serious problem with the extension: the central charge
becomes infinite. 
Formally, 
\bes
J_X(z)J_Y(w) &\sim& \frac1{z-w} J_{[X,Y]}(w) \\
&\mp& \frac{1}{(z-w)^2} \iint \dx\dy \tr(X(x,z)Y(y,w))\, \delta(x-y)\delta(y-x).
\eens
The double Wick contraction is proportional to $\delta(0)$, since
\be
\delta(x-y) \delta(y-x) = \delta(0) \delta(x-y),
\ee
and hence the central charge is infinite,
\be
k = \mp y_M \delta(0).
\ee
Clearly, an infinite central charge does not make sense. The generalization
of the affine algebra to higher dimensions must be done differently, a
task to which we now turn.

\subsection{ $p$-jet quantization }

To overcome the problem with the infinite central charge, we replace fields with
the corresponding $p$-jets. To maintain as much as possible of the field
formalism, the Taylor coefficients are not displayed explicitly, but instead 
everything is expressed in terms of the generating functions (\ref{phiR}) and
(\ref{piR}). The OPE of the fundamental fields remains essentially unchanged;
the only difference compared to (\ref{ope2}) is that the delta function has been
replaced by its truncation $\delta_p(x,y)$,
\bes
\phi(x,z)\pi(y,w) &\sim& \frac{1}{z-w}\, \delta_p(x,y), 
\nlb{ope3}
\pi(x,z)\phi(y,w) &\sim& \frac{\mp1}{z-w}\, \delta_p(y,x)
\eens
Recall that the $p$-jet delta function is not symmetric.
Because we deal with relative fields,
\be
X(x,z) = X(x+q(z)),
\ee
so the smeared generators (\ref{JXzabs}) are replaced by
\be
J_X(z) = \int \dx \no{ \pi(x,z) X(x+q(z)) \phi(x,z) }.
\label{JXz}
\ee
To see that this expression is indeed equal to the gauge algebra generators in
\cite{Lar98}, we use the definitions (\ref{phiR}) and (\ref{piR}) and
suppress the $z$ dependence. First integrate repeatedly by parts:
\bes
J_X &=& \int : 
 \Big(\summp (-)^{|\mm|} \pi^\mm \d_\mm\delta(x)\Big)
 \Big(\sum_\rr \frac1{\rr!} \d_\rr X(q) x^\rr \Big)
 \Big(\sumnp \frac1{\nn!}\phi_\nn x^\nn\Big) : 
\nl
&=& \sum\sum\sum \frac1{\nn!\rr!} \no{\pi^\mm \d_\rr X(q) \phi_\nn}
 \int \dx \d_\mm (x^{\nn+\rr}) \delta(x).
\ees
The last integral equals
$\mm!\, \delta^{\nn+\rr}_\mm$,
and hence
\be
J_X = \summp \sumnp \pi^\mm J^\nn_\mm(X(q)) \phi_\mm,
\label{JXjet}
\ee
where
\be
J^\nn_\mm(X) = \binom{\mm}{\nn}\d_{\mm-\nn} X.
\label{Jnm}
\ee
Since the binomial coefficients vanish whenever $\nn<\mm$, the sum over $\nn$ is
in fact restricted to this range.
Equations (\ref{JXjet}) and (\ref{Jnm}) are thus equal to equations (6.6) and (6.4)
of \cite{Lar98}, respectively.

The OPE $J_X(z) J_Y(w)$ becomes
\bes
&& \frac1{z-w} \iint \dx\dy \delta_p(x,y) \no{ \pi(x,z) X(x,z)Y(y,w)\phi(y,w) } \nl
&-& \frac1{z-w} \iint \dx\dy \delta_p(y,x) \no{ \pi(y,w) Y(y,w) X(x,z) \phi(x,z) } \nl
&\mp& \frac1{(z-w)^2} \iint \dx\dy \delta_p(x,y)\delta_p(y,x)\, \tr(X(x,z) Y(y,w)). \nl
\label{JJ-ope}
\ees
We must be careful when evaluating this expression, because 
$\delta_p(x,y) \neq \delta_p(y,x)$.
Suppressing normal ordering and the $z$ and $w$ dependence, the inner integral 
in the first term can be rewritten as
\bes
&& \int \dy \delta_p(x,y) \pi(x) X(x)Y(y)\phi(y) \nl
&&=\, \pi(x) X(x) (Y(x)\phi(x))\rvert_p \nle
&&=\, \pi(x) (X(x) (Y(x)\phi(x))\rvert_p )\rvert_p \nl
&&=\, \pi(x) (X(x) Y(x)\phi(x) )\rvert_p,
\eens
where we used the properties (\ref{piprop}) and (\ref{pjetprop}). 
Finally using (\ref{piprop}) once again, the first term in (\ref{JJ-ope}) becomes
\be
\frac1{z-w} \int \dx \no{ \pi(x,z)X(x,z)Y(x,z)\phi(x,z) }
\ee
The double Wick contraction is calculated using (\ref{sum1}), and equals
\bes
&&\frac{\mp1}{(z-w)^2} \int \dx\dy A_{d,p} \delta(x)\delta(y)\, \tr(X(x,z) Y(y,w)) 
\nle
&=& \frac{\mp1}{(z-w)^2} \binom{d+p}{d}\, \tr(X(0,z) Y(0,w)),
\eens
where we used (\ref{Adp}) in the last step.

Hence the OPE is
\bes
J_X(z) J_Y(w) &\sim& \frac1{z-w} J_{[X,Y]}(w) 
\nle
  &&\mp\ \frac1{(z-w)^2} \binom{d+p}{d} \tr(X(0,z) Y(0,w)).
\eens
Equivalently, the operators
\be
\J_X = \frac1{2\pi i}\oint dz\, J_X(z)
\label{JX}
\ee
satisfy the Lie algebra brackets
\bes
[\J_X, \J_Y] &=& \frac{-1}{4\pi}\oint_0 dw\, \oint_w dz\, J_X(z)J_Y(w)
\nle
&=& \J_{[X,Y]} + \frac{k}{2\pi i} \oint dw\, \dot X^a(0,w) Y^a(0,w).
\eens
where
\be
k = \mp \binom{d+p}{d} y_M,
\label{k}
\ee
and the dot denotes the partial derivative w.r.t. the complex coordinate,
$\dot f(x,z) \equiv \d f(x,z)/\d z$.
In particular, with $X(x,z) = X(x+q(z))$, 
\be
[\J_X, \J_Y] = \J_{[X,Y]} 
 + \frac{k}{2\pi i} \oint dw\, \dot q^\mu(w) \dmu X^a(q(w)) Y^a(q(w)),
\label{Affd}
\ee
which is the form of $Aff(d,\oj)$ described in \cite{Lar98}.

Note that this cocycle is proportional to the second Casimir operator.
$Aff(d,\oj)$ is thus unrelated to the gauge anomalies appearing in the
standard model of particle physics, which are proportional to the third Casimir.

In one dimension, the cocycle can be rewritten as
$\int dq\, X_a'(q)Y^a(q)$, 
which shows that (\ref{Affd}) reduces to the affine Kac-Moody algebra
$\widehat \oj$ when $d=1$.

\section{ Multi-dimensional Virasoro algebra }

\subsection{ Classical representations }

Let $\xi=\xmu(x)\dmu$ be a vector field, with commutator
\be
[\xi,\eta] \equiv \xmu\dmu\ynu\dnu - \ynu\dnu\xmu\dmu.
\ee
The algebra of vector fields in $d$ dimensions, denoted by $\vect(d)$,
is generated by the Lie derivatives $\L_\xi$. The bracket reads
\be
[\L_\xi,\L_\eta] &=& \L_{[\xi,\eta]}.
\label{vectd}
\ee
We will colloquially refer to $\vect(d)$ as the $d$-dimensional 
diffeomorphism algebra, although it is not strictly the Lie algebra of
the diffeomorphism group in $d$ dimensions.

The most natural type of $\vect(d)$ representations act on modules of
tensor densitities. In one dimension, these are the primary fields of CFT.
Consider the Heisenberg algebra generated by a tensor-valued spacetime field 
$\phi(x)$ and its canonical momentum $\pi(x)$:
\be
[\phi(x), \pi(y)] = i\delta(x-y), \qquad
[\phi(x), \phi(y)] = [\pi(x), \pi(y)] = 0.
\ee
$\vect(d)$ can be embedded into this Heisenberg algebra as follows:
\be
\L_\xi = -i \int \dx \Big( \xmu(x)\pi(x)\dmu\phi(x) 
 + \dnu\xmu(x) \pi(x) T^\nu_\mu \phi(x) \Big),
\label{LphiA}
\ee
where the matrices $T^\mu_\nu$ satisfy $\gl(d)$:
\be
[T^\mu_\rho, T^\nu_\si] = \delta^\nu_\rho T^\mu_\si - \delta^\mu_\si T^\nu_\rho.
\ee
For every $gl(d)$ representation $\varrho$, the embedding (\ref{LphiA}) yields a $\vect(d)$
representation acting on tensor densitities of type $\varrho$:
\be
[\L_\xi, \phi(x) ] = -\xmu(x)\dmu\phi(x) - \dnu\xmu(x) T^\nu_\mu \phi(x).
\ee
The conjugate momentum transforms as a density in the dual representation:
\be
[\L_\xi, \pi(x) ] = -\xmu(x)\dmu\pi(x) -\dmu\xmu(x)\pi(x) 
 + \dnu\xmu(x) \pi(x) T^\nu_\mu.
\ee

\subsection{ Relative fields }

The fields in the classical embedding (\ref{LphiA}) are absolute fields, 
To prepare for quantization, we follow the same recipe as for gauge algebras;
replace absolute fields with relative fields and shift 
$x^\mu \to x^\mu + q^\mu$ to make the dependence on the
observer's position manifest. This leads to
\be
\L^\phi_\xi = -i \int \dx \Big( \xmu(x+q) \pi(x)\dmu\phi(x) 
 + \dnu\xmu(x+q) \pi(x)T^\nu_\mu \phi(x) \Big),
\label{LphiR}
\ee
However, diffeomorphisms do not commute with the observer's position. Hence we also 
need a second class of $\vect(d)$ representations, acting nonlinearly on $q^\mu$.
The embedding of $\vect(d)$ is given by
\be
\L^q_\xi = i \xmu(q) p_\mu,
\label{Lq}
\ee
where the $2d$ generators $q^\mu$ and $p_\nu$ satisfy the CCR (\ref{H-q-p}).
The action on the observer's position is non-linear:
\bes
[\L^q_\xi, q^\mu] &=& \xmu(q), \nle
{[}\L^q_\xi, p_\nu] &=& -\dnu\xmu(q)\dmu.
\eens
These relations are to be understood as defining nonlinear realizations in 
the space of polynomial functions of $q^\mu$, 
If $\Phi(q)$ is such a function, the operators $\L^q_\xi$ act as scalar fields:
\be
[\L^q_\xi, \Phi(q)] = \xmu(q) \dmu\Phi(q).
\ee

The operators in (\ref{LphiR}) and (\ref{Lq}) both satisfy $\vect(d)$ individually, but 
their sum $\L^\phi_\xi + \L^q_\xi$ does not, because the observer's momentum 
does not commute with $\xmu(x+q)$.
To remedy this defect, we introduce the improved momentum
\be
P_\mu = p_\mu + \int \dx \pi(x) \dmu \phi(x),
\label{Pmu}
\ee
which satisfies
\bes
[P_\mu, q^\nu] &=& -i\delta^\nu_\mu, \nl
{[}P_\mu, \phi(x)] &=& -i\d_\mu\phi(x), \nle
{[}P_\mu, \pi(x)] &=& -i\d_\mu\pi(x), \nl
{[}P_\mu, P_\nu] &=& 0.
\eens
The improved generators
\be
\L^{\prime q}_\xi = \xmu(q)P_\mu
\ee
satisfy $[\L^{\prime q}_\xi, \L^{\prime q}_\eta] = \L^{\prime q}_{[\xi,\eta]}$
and $[\L^{\prime q}_\xi,\L^\phi_\eta] = 0$. The sum
\be
\L_\xi = \L^{\prime q}_\xi + \L^\phi_\xi
\label{Lxi0}
\ee
hence furnishes a realization of $\vect(d)$ which acts correctly on both the observer's
position and on the relative fields.

\subsection {Quantization }

The OPEs between the fundamental fields are defined to be
\bes
q^\mu(z) p_\nu(w) &\sim& \frac{1}{z-w}\, \delta^\mu_\nu, 
\nle
\phi(x,z)\pi(y,w) &\sim& \frac{1}{z-w}\, \delta_p(x,y), 
\eens
whereas all other contractions vanish.

It is useful to arrange the diffeomorphism generators somewhat 
differently than in (\ref{Lxi0}). Set
\bes
L^0_\xi(z) &=& -\no{ \xmu(q(z)) p_\mu(z) } \nl
L^1_\xi(z) &=& \int \dx (\xmu(x+q(z)) - \xmu(q(z))) \no{\pi(x,z)\dmu\phi(x,z) }, \nl
L^2_\xi(z) &=& \int \dx \dnu\xmu(x+q(z)) \no{ \pi(x,z) T^\nu_\mu\phi(x,z) }.
\label{Lixi}
\ees
The total $Vir(d)$ generators are
\be
L_\xi(z) = L^0_\xi(z) + L^1_\xi(z) + L^2_\xi(z).
\label{Lxiz}
\ee

The OPEs with the fundamental fields are
\bes
L_\xi(z) q^\mu(w) &\sim& \frac1{z-w}\, \xmu(q(w)), \nl
L_\xi(z) p_\nu(w) &\sim& \frac1{z-w} \Big(
 -\dnu\xmu(q(w)) p_\mu(w) \nl
 &+& \int \dx (\dnu\xmu(x+q(w)) - \dnu\xmu(q(w))) \no{\pi(x,w)\dmu\phi(x,w) } \nl
 &+& \int \dx \dnu\drho\xmu(x+q(w)) \no{ \pi(x,w) T^\rho_\mu\phi(x,w) }
 \Big), \nle
L_\xi(z)\phi(x,w) &\sim& \frac1{z-w} \Big(
 -(\xmu(x+q(w)) - \xmu(q(w))) \dmu \phi(x,w) \nl
 &-& \dnu\xmu(x+q(w)) T^\nu_\mu \phi(x,w),
 \Big)
\nl
L_\xi(z)\pi(x,w) &\sim& \frac1{z-w}\Big(
 -(\xmu(x+q(w)) - \xmu(q(w))) \dmu \pi(x,w) \nl
 &+& \dnu\xmu(x+q(w)) \pi(x,w) T^\nu_\mu,
 \Big).
\eens
To calculate the OPE $L_\xi(z) L_\eta(w)$ is quite tedious, and is deferred to
Appendix \ref{app:LL}. The result is
\be
L_\xi(z) L_\eta(w) &\sim& \frac{L_{[\xi,\eta]}(w)}{z-w}
+ \frac{Z_\xeta(z,w)}{(z-w)^2},
\label{LLzw}
\ee
where
\bes
Z_\xeta(z,w) &=& -\ c_1\, \dnu\xmu(q(z))\, \dmu\ynu(q(w))
 \nlb{Zxieta}
 &&-\ c_2\, \dmu\xmu(q(z))\, \dnu\ynu(q(w)).
\eens
Assume that the fields transform in a representation $\varrho$ of
$\gl(d)$, and let I be the unit matrix in this representation.
\bes
\tr\,I &=& \dim\, \varrho\, \equiv\, \Delta_\varrho, \nl
\tr\,T^\mu_\nu &=& k_0(\varrho) \delta^\mu_\nu, 
\label{glparam} \\
\tr\,T^\mu_\rho T^\nu_\si &=& 
 k_1(\varrho) \delta^\mu_\rho \delta^\nu_\si +
 k_2(\varrho) \delta^\mu_\si \delta^\nu_\rho.
\eens
The {\em abelian charges} $c_1$ and $c_2$ are given by
\bes
c_1 &=& 1 \pm \Big\{ \binom{d+p+1}{d+2} \Delta_\varrho 
 + \binom{d+p}{d} k_1(\varrho) \Big\}, 
\nlb{c1c2}
c_2 &=& \pm \Big\{ \binom{d+p}{d+2} \Delta_\varrho 
 + 2\binom{d+p}{d+1} k_0(\varrho) 
 + \binom{d+p}{d} k_2(\varrho) \Big\},
\eens
where the upper sign refers to bosonic fields and the lower to fermionic fields. 

The $\gl(d)$ parameters (\ref{glparam}) can be rewritten using more conventional
notation if we note that $\gl(d) = \ssl(d) \oplus \gl(1)$. A $\gl(d)$ matrix is
of the form
\be
T^\mu_\nu = S^\mu_\nu + \kappa \delta^\mu_\nu I,
\label{Tmunu}
\ee
where $S^\mu_\nu$ is an $\ssl(d)$ matrix and
$\kappa$ is the weight of the field $\phi(x,z)$ as a density.
By definition,
\be
S^\mu_\mu = 0.
\ee
$\gl(d)$ representations are labelled by a pair $(\varrho,\kappa)$, where
$\varrho$ now is an $\ssl(d)$ representation.
The $\ssl(d)$ traces are
\bes
\tr\,S^\mu_\nu &=& 0, 
\nle
\tr\,S^\mu_\rho S^\nu_\si &=& 
 y_\varrho (\delta^\mu_\si\delta^\nu_\rho - \frac1{d}\delta^\mu_\rho \delta^\nu_\si),
\eens
where $y_\varrho$ is the value of the quadratic Casimir in $\varrho$.
The last condition guarantees that $\tr\,S^\mu_\mu S^\nu_\si = 0$.
The $\gl(d)$ traces can now be written as
\bes
\tr\,I &=& \Delta_\varrho, \nl
\tr\,T^\mu_\nu &=& \kappa \Delta_\varrho \delta^\mu_\nu, \\
\tr\,T^\mu_\rho T^\nu_\si &=& 
 y_\varrho \delta^\mu_\si\delta^\nu_\rho +
 (\kappa^2\Delta_\varrho - \frac{y_\varrho}{d}) \delta^\mu_\rho \delta^\nu_\si.
\eens
Hence
\bes
k_0(\varrho) &=& \kappa \Delta_\varrho, \nl
k_1(\varrho) &=& y_\varrho, \\
k_2(\varrho) &=& \kappa^2\Delta_\varrho - \frac{y_\varrho}{d}.
\eens

Finally we express the OPE (\ref{LLzw}) -- (\ref{Zxieta}) as a Lie algebra.
Define
\be
\L_\xi = \frac1{2\pi i} \oint dz\, L_\xi(z).
\label{Lxi}
\ee
These operators satisfy the {\em multi-dimensional Virasoro algebra} $Vir(d)$, i.e.
\bes
[\L_\xi, \L_\eta] &=& \L_{[\xi,\eta]} + \Z_\xeta, \nl
{[}\L_\xi, q^\mu(t)] &=& \xmu(q(t)),
\label{Vird}\\
{[}q^\mu(t), q^\nu(t')] &=& 0.
\eens
where the extension is
\bes
\Z_\xeta &=& \frac{-1}{2\pi i} \oint dz\, \Big(
 c_1\, \dnu\dot\xi^\mu(q(z))\, \dmu\ynu(q(z)) + 
 c_2\, \dmu\dot\xi^\mu(q(z))\, \dnu\ynu(q(z)) 
 \Big) \nl
&=& \frac{-1}{2\pi i} \oint dz\, \dot q^\rho(z) \Big(
 c_1\, \drho\dnu\xmu(q(z))\, \dmu\ynu(q(z)) 
 \\
 &&\qquad +\ c_2\, \drho\dmu\xmu(q(z))\, \dnu\ynu(q(z))
 \Big).
\eens
The cocycle proportional to $c_1$ was discovered by Rao and Moody
\cite{RM94}, and the one proportional to $c_2$ by myself \cite{Lar91}.

\subsection{ Intertwining action on $Aff(d,\oj)$ }
\label{ssec:VirAff}

In this subsection we complete the full semi-direct product
$Vir(d) \ltimes Aff(d,\oj)$, or more precisely the extension of 
$\vect(d) \ltimes \map(d,\oj)$. There is a distinction because the
mixed bracket also acquires an extension, although it vanishes in the
case that $\oj$ is semisimple.

The OPE between the operators (\ref{Lxiz}) and (\ref{JXz}) reads
\be
L_\xi(z) J_X(w) \sim \frac{J_{\xi X}(w)}{z-w} + \frac{W_\xX(z,w)}{(z-w)^2},
\label{LJzw}
\ee
where $X(z)$ transforms as a density of weight one:
\be
\xi X = \xmu\dmu X + \dmu\xmu X.
\ee
The extension is non-zero only if the trace does not vanish in the $\oj$
representation $M$. Assume that the $\oj$ matrices satisfy
\be
\tr\,M^a = z_M\delta^a,
\ee
where $\delta^a$ is a privileged matrix in $M$. The architypical case is
$\oj = \gl(1)$, where the privileged matrix is unity. If the
parameter $z_M$ is nonzero, the extension in (\ref{LJzw}) becomes
\be 
W_\xX(z,w) = c_7\, \dmu\xmu(q(z))\, \delta^a X^a(q(w)).
\label{WxX}
\ee
The value of the abelian charge $c_7$ equals
\be
c_7 = \mp z_M \Big\{ \binom{d+p}{d+1} \Delta_\varrho + \binom{d+p}{d} k_0(\varrho) \Big\},
\label{c7}
\ee
where $\Delta_\varrho$ and $k_0(\varrho)$ were defined in (\ref{glparam}).

The OPE (\ref{LJzw}) corresponds to the following bracket between 
$\L_\xi$ and $\J_X$ (\ref{Lxi}) and (\ref{JX}):
\bes
[\L_\xi, \J_X] &=&
 \J_{\xi X} + \frac{c_7}{2\pi i} \oint dz\, \dmu\dot\xi^\mu(q(z))\, \delta^a X^a(q(z)) 
\\
&=& \J_{\xi X} + \frac{c_7}{2\pi i} \oint dz\, \dot q^\rho(z) \drho\dmu\xmu(q(z))\, \delta^aX^a(q(z)).
\eens
Finally we must correct the previously computed abelian charges for the fact that
$M$ commutes with $\vect(d)$ and $\varrho$ commutes with $\map(d,\oj)$.
The fields have hence additional indices on which the various algebras act trivially,
each of which contributes an equal amount to the abelian charges.
The $Vir(d)$ charges $c_1$ and $c_2$ in (\ref{c1c2}) are multiplied with
$\Delta_M$, and and $Aff(d,\oj)$ charge $k = c_5$ in (\ref{k}) is multiplied with
$\Delta_\varrho$. 
This correction is accounted for in the summary in section \ref{sec:summary}.

The proof of the formulas in this section is given in Appendix \ref{app:VirAff}.

\section{ Reparametrization Virasoro algebra }
\label{sec:repar}

The fields $\phi(x,z)$ depend not only on the spacetime coordinate $x$ but
also on the holomorphic variable $z$, which was identified as a time-like 
parameter along the observer's spacetime trajectory in \cite{Lar98}.
We can therefore extend an $Vir(d)\ltimes Aff(d,\oj)$ representation
to a representation of reparametrizations ``for free'', i.e. without
introducing new field components. After quantization, reparametrizations
generate a Virasoro algebra which intertwines with the multi-dimensional
Virasoro and affine algebras.

All proofs in this section are carried out in Appendix \ref{app:repar}.

\subsection{ The energy-momentum tensor in CFT }

The energy-momentum of a conformal field $\phi(z)$ and its conjugate
momentum $\pi(z)$ is
\be
T(z) = -\no{\pi(z)\dot\phi(z)} + \lambda \frac{d}{dz}(\no{\pi(z)\phi(z)}),
\label{EMT}
\ee
where $\lambda$ is the conformal weight of $\phi(z)$, not to be confused with the
weight $\kappa$ as a tensor density (\ref{Tmunu}). The conjugate momentum
$\pi(z)$ has conformal weight $1-\lambda$.

The OPEs between $T(z)$ and the fundamental fields are
\bes
T(z)\phi(w) &\sim& \frac{\dot\phi(w)}{z-w} + \frac{\lambda}{(z-w)^2} \phi(w), 
\nle
T(z)\pi(w) &\sim& \frac{\dot\pi(w)}{z-w} + \frac{1-\lambda}{(z-w)^2} \pi(w).
\eens
In particular, if $\phi(z)$ has conformal weight $\lambda=1$,
\bes
T(z) &=& \no{\dot\pi(z)\phi(z)}, 
\nlb{weight1}
T(z)\phi(w) &\sim& \frac{d}{dw} \Big( \frac{\phi(w)}{z-w} \Big).
\eens
The OPE of the energy-momentum tensor with itself reads
\be
T(z)T(w) &\sim& \frac{\dot T(w)}{z-w} + \frac{2T(w)}{(z-w)^2} 
 + \frac{c/2}{(z-w)^4},
\label{TT}
\ee
where $c$ is the central charge.
\be
c = \pm 2 (6\lambda^2 - 6\lambda + 1),
\label{clambda}
\ee
where as usual the upper (lower) sign applies to bosonic (fermionic) fields.

\subsection{ Reparametrization algebra }

Reparametrizations act on two types of fields: the observer's trajectory
$q^\mu(z)$ and the $p$-jet fields $\phi(x,z)$, and also on their
canonical momenta. The definition (\ref{EMT}) must therefore be replaced by
\bes
T(z) &=& -\no{\dot q^\mu(z)p_\mu(z)} \\
 &&+\ \int \dx \Big( -\no{\pi(x,z)\dot\phi(x,z)} 
   + \lambda \frac{d}{dz}(\no{\pi(x,z)\phi(x,z)}) \Big),
\eens
The OPEs between the reparametrization generators and 
the fundamental fields read:
\bes
T(z) q^\mu(w) &\sim& \frac{\dot q^\mu(w)}{z-w}, \nl
T(z) p_\nu(w) &\sim& \frac{d}{dw} \Big( \frac{p_\nu(w)}{z-w} \Big),
\nle
T(z)\phi(x,w) &\sim& \frac{\dot\phi(x,w)}{z-w} + \frac{\lambda}{(z-w)^2} \phi(x,w), \nl
T(z)\pi(x,w) &\sim& \frac{\dot\pi(x,w)}{z-w} + \frac{1-\lambda}{(z-w)^2} \pi(x,w).
\eens
The OPE $T(z)T(w)$ is given by (\ref{TT}), and the central charge equals
\be
c \equiv c_4 = 2d\, \pm\, 2 (6\lambda^2 - 6\lambda + 1)\binom{d+p}{d} \Delta_\varrho \Delta_M.
\label{c4}
\ee
We recognize the contributions from the $2d$ fields $q^\mu(z)$ and $p_\nu(z)$, and
from the $\binom{d+p}{d} \Delta_\varrho\Delta_M$ field pairs $\phi(x,z)$ and $\pi(x,z)$.
The smeared operators 
\be
\T_f = -\frac1{2\pi i}\oint dz\, f(z)T(z)
\ee
generate the Virasoro algebra $Vir(1)$:
\be
[\T_f,\T_g] = \T_{[f,g]} 
 - \frac{c_4}{24\pi i}\oint dz\, \ddot f(z)\dot g(z),
\ee
where $[f,g] = f\dot g - g\dot f$.

\subsection{ Intertwining action on $Vir(d)$ and $Aff(d,\oj)$ }

Conformal weights are additive. If the fields $\phi_1(x,z)$ and $\phi_2(x,z)$
have weights $\lambda_1$ and $\lambda_2$, the product $\phi_1(x,z)\phi_2(x,z)$
has weight $\lambda_1+\lambda_2$. In particular, since the observer's 
trajectory $q^\mu(z)$ has zero weight, so has any function $X(q(z))$.
Hence the diffeomorphism and gauge generators (\ref{Lxiz}) and (\ref{JXz})
have weight $(1-\lambda) + 0 + \lambda = 1$. 
The OPEs between the reparametrization algebra $Vir(1)$ and $Vir(d)$ and 
$Aff(d,\oj)$ are
\bes
T(z)L_\xi(w) &\sim& \frac{d}{dw} \Big( \frac{L_\xi(w)}{z-w} \Big) 
 + \frac{c_3}{(z-w)^3} \dmu\xmu(q(w)),
\label{TL-ope}\\
T(z)J_X(w) &\sim& \frac{d}{dw} \Big( \frac{J_X(w)}{z-w} \Big) 
 + \frac{c_6}{(z-w)^3} \delta^a X^a(q(w)),
\label{TJ-ope} 
\ees
where two new abelian charges were introduced:
\bes
c_3 &=& 1\, \pm\, (2\lambda-1) \Delta_M \Big\{ \binom{d+p}{d+1}\Delta_\varrho
 + \binom{d+p}{d}k_0(\varrho) \Big\},
\label{c3}\\
c_6 &=& \pm (2\lambda-1) z_M \binom{d+p}{d} \Delta_\varrho.
\label{c6} 
\ees
Note the contribution from the observer's position to $c_3$.
The corresponding Lie brackets between smearing operators are
\bes
[\T_f, \L_\xi] &=& -\frac{c_3}{4\pi i} \oint dz\, \ddot f(z) \dmu\xmu(q(z)), 
\nle
[\T_f, \J_X] &=& -\frac{c_6}{4\pi i} \oint dz\, \ddot f(z) \delta^a X^a(q(z)).
\eens

\section{ Summary of the main formulas }
\label{sec:summary}

In this section we collect the main equations for easy reference.

\smallskip\noindent
Operator product expansions:
\bes
L_\xi(z) L_\eta(w) &\sim& \frac{L_{[\xi,\eta]}(w)}{z-w} 
- \frac{1}{(z-w)^2} \Big\{
 c_1\, \dnu\xmu(q(z))\, \dmu\ynu(q(w)) \nl
 &&\quad+\ c_2\, \dmu\xmu(q(z))\, \dnu\ynu(q(w))
 \Big\},
\nl
T(z)L_\xi(w) &\sim& \frac{d}{dw} \Big( \frac{L_\xi(w)}{z-w} \Big)
 + \frac{c_3}{(z-w)^3} \dmu\xmu(q(w)),
\nl
T(z)T(w) &\sim& \frac{\dot T(w)}{z-w} + \frac{2T(w)}{(z-w)^2}
 + \frac{c_4/2}{(z-w)^4},
\nl
J_X(z)J_Y(w) &\sim& \frac{J_{[X,Y]}(w)}{z-w} 
+ \frac{1}{(z-w)^2} \Big\{ c_5\, X^a(q(z))\, Y^a(q(w)) \nl
&&\quad+\ c_8\, \delta^a X^a(q(z))\, \delta^b Y^b(q(w)) \Big\},
\nle
T(z)J_X(w) &\sim& \frac{d}{dw} \Big( \frac{J_X(w)}{z-w} \Big)
 + \frac{c_6}{(z-w)^3} \delta^a X^a(q(w)),
\nl
L_\xi(z) J_X(w) &\sim& \frac{J_{\xi X}(w)}{z-w}
+ \frac{c_7}{(z-w)^2} \dmu\xmu(q(z))\, \delta^a X^a(q(w)),
\nl
L_\xi(z) q^\mu(w) &\sim& \frac1{z-w}\, \xmu(q(w)),
\nl
J_X(z) q^\mu(w) &\sim& 0,
\nl
T(z) q^\mu(w) &\sim& \frac{\dot q^\mu(w)}{z-w},
\nl
q^\mu(z) q^\nu(w) &\sim& 0.
\eens

\noindent
Brackets:
\bes
{[}\xi,\eta] &=& \xmu\dmu\ynu\dnu - \ynu\dnu\xmu\dmu, \nl
{[}X,Y] &=& i f^{abc} X^a Y^b J^c, \nle
{[}f,g] &=& f\dot g - g\dot f, \nl
\xi X &=& \xmu\dmu X + \dmu\xmu X.
\eens

\noindent
Traces in the $\gl(d)\oplus \oj$ representation $\varrho\oplus M$:
\bes
\tr_\varrho(I) &=& \Delta_\varrho, \nl
\tr_\varrho(T^\mu_\nu) &=& k_0(\varrho), \nl
\tr_\varrho(T^\mu_\rho T^\nu_\si) &=& k_1(\varrho) \delta^\mu_\si \delta^\nu_\rho
 + k_2(\varrho) \delta^\mu_\rho \delta^\nu_\si, \\
\tr_M(I) &=& \Delta_M, \nl
\tr_M(M^a) &=& z_M \delta^a, \nl
\tr_M(M^a M^b) &=& y_M \delta^{ab} + w_M \delta^a \delta^b.
\eens
Note the additional term proportional to $w_M$ added to the last equation.
Such a term is possible in general, but vanishes if $\oj$ is semisimple.

\smallskip\noindent 
Abelian charges:
\bes
c_1 &=& 1 \pm \Delta_M \Big\{ \binom{d+p+1}{d+2} \Delta_\varrho
 + \binom{d+p}{d} k_1(\varrho) \Big\},
\nl
c_2 &=& \pm \Delta_M \Big\{ \binom{d+p}{d+2} \Delta_\varrho
 + 2\binom{d+p}{d+1} k_0(\varrho)
 + \binom{d+p}{d} k_2(\varrho) \Big\},
\nl
c_3 &=& 1\, \pm\, (2\lambda-1) \Delta_M \Big\{ \binom{d+p}{d+1}\Delta_\varrho
 + \binom{d+p}{d}k_0(\varrho) \Big\}
\nl
c_4 &=& 2d\, \pm\, 2 (6\lambda^2 - 6\lambda + 1)\binom{d+p}{d} \Delta_\varrho \Delta_M,
\nle
c_5 &=& \mp \binom{d+p}{d} y_M \Delta_\varrho,
\nl
c_6 &=& \pm (2\lambda-1) z_M \binom{d+p}{d}\Delta_\varrho.
\nl
c_7 &=& \mp z_M \Big\{ \binom{d+p}{d+1} \Delta_\varrho + \binom{d+p}{d} k_0(\varrho) \Big\},
\nl
c_8 &=& \mp w_M \binom{d+p}{d} \Delta_\varrho.
\eens

\noindent 
Lie algebra generators:
\bes
\L_\xi &=& \frac1{2\pi i} \oint dz\, L_\xi(z)
\nl
\J_X &=& \frac1{2\pi i} \oint dz\, J_X(z)
\\
\T_f &=& -\frac{1}{2\pi i}\oint dz\, f(z)T(z)
\eens

\noindent
Lie brackets:
\bes
{[}\L_\xi, \L_\eta] &=& \L_{[\xi,\eta]}
 - \frac1{2\pi i} \oint dz\, \dot q^\rho(z) \Big(
 c_1\, \drho\dnu\xmu(q(z))\, \dmu\ynu(q(z)) \nl
&&\qquad +\ c_2\, \drho\dmu\xmu(q(z))\, \dnu\ynu(q(z)) \Big),
\nl
{[}\T_f, \L_\xi] &=& -\frac{c_3}{4\pi i} \oint dz\, \ddot f(z) \dmu\xmu(q(z)),
\nl
{[}\T_f,\T_g] &=& \T_{[f,g]}
 - \frac{c_4}{24\pi i}\oint dz\, \ddot f(z)\dot g(z), \nl
\nl
{[}\J_X, \J_Y] &=& \J_{[X,Y]}
 + \frac1{2\pi i} \oint dz\, \dot q^\rho(z) \Big( c_5\, \drho X^a(q(z))\, Y^a(q(z)) \nl
 &&\qquad +\ c_8\, \delta^a \drho X^a(q(z))\, \delta^b Y^b(q(z)) \Big),
\nle
{[}\T_f, \J_X] &=& -\frac{c_6}{4\pi i} \oint dz\, \ddot f(z) \delta^a X^a(q(z)),
\nl
{[}\L_\xi, \J_X] &=& \J_{\xi X}
 + \frac{c_7}{2\pi i} \oint dz\, \dot q^\rho(z) \drho\dmu\xmu(q(z))\, \delta^aX^a(q(z)),
\nl
\nl
{[}\L_\xi, q^\mu(z)] &=& \xmu(q(z)),
\nl
{[}\J_X, q^\mu(z)] &=& 0,
\nl
{[}\T_f, q^\mu(z)] &=& -f(z) \dot q^\mu(z),
\nl
{[}q^\mu(z), q^\nu(w)] &=& 0.
\eens

The abelian charges agree with the result in \cite{Lar98}, equations (5.6) and (6.5).
The Lie brackets are the same as in equations (2.3) and (6.2) of that paper,
apart from some cohomologically trivial terms,
but all extensions have the opposite signs. The reason is that the standard OPE 
formalism used in the present article is based on highest-weight representations, 
whereas in \cite{Lar98} we instead considered representations of 
lowest-energy type. In physics the latter are more natural -- energy
is typically bounded from below rather than from above -- but the highest-weight
convention is standard in CFT, and we wanted to facilitate comparison with the standard
reference \cite{FMS94}.

\section{ Conclusion }

The technical results in this paper are identical to those in \cite{Lar98},
but the presentation is quite different and hopefully more accessible. 
One improvement is that the Taylor coefficients have been hidden in
generating functions, thus making the equations more similar to the 
analogous formulas for fields. That we really deal with $p$-jets is manifested
by the replacement of delta functions by their $p$-jet counterparts, which
can be multiplied with each other in a meaningful way.

The second improvement is the use of the standard OPE formalism to calculate
brackets and extensions. The formalism in \cite{Lar98}, albeit correct, was
somewhat non-standard and cumbersome. The OPE formalism used here is simpler
and more standard.

The obvious physical application of the multi-dimensional Virasoro algebra,
in particular $Vir(4)$, is in quantum gravity. In fact, it is a necessary 
ingredient in any local quantum theory of gravity. Recall the standard argument
why there can be no local observables in quantum gravity:
\begin{enumerate}
\item
In quantum theory an observable is a gauge-invariant operator.
\item
In general relativity, all spacetime diffeomorphisms are gauge symmetries.
\item
Hence an observable in quantum gravity must commute with all diffeomorphisms,
i.e. it can not depend on local coordinates.
\end{enumerate}

The weak link in this argument is that in order to go from 2 to 3, one must
assume that classical and quantum gravity have the same gauge symmetries. If
the diffeomorphism algebra acquires an extension upon quantization, this assumption 
is false. $Vir(d)$ is the extension of $\vect(d)$ which is necessary for
local observables, just as a non-zero central charge is necessary for local
observables in CFT. 
That a diffeomorphism algebra extension is necessary for locality was
emphasized in \cite{Lar14}.

\section{Appendices}

\appendix

\section{Proof of the smearing properties (\ref{delta_smear}) -- (\ref{partint_y})}
\label{app:pjet}

\noindent 
Integrate by parts repeatedly.
\bes
\int \dy f(y) \delta_p(x,y)
&=& \summp \mfrac\, x^\mm \int \dy f(y) \d_\mm\delta(y) \nl
&=& \summp \frac1{\mm!}\, x^\mm \int \dy \d_\mm f(y)\, \delta(y) 
\nl
&=& \summp \frac1{\mm!}\, \d_\mm f(0)\, x^\mm
= f(x) |_p
\eens
\bes
\int \dy f(y) \d^x_\mu\delta_p(x,y) 
&=& \summp \mfrac\, m_\mu x^{\mm-\mu} \int \dy f(y)\, \d_\mm\delta(y) \nl
&=& \summp \frac1{(\mm-\mu)!}\, x^\mm \int \dy \d_\mm f(y)\, \delta(y) 
\nl
&=& \summp \frac1{\nn!}\, \d_{\nn+\mu} f(0)\, x^\nn
= \dmu f(x) |_p
\eens
\bes
\int \dy f(y) \d^y_\mu\delta_p(x,y)
&=& \summp \mfrac\, x^\mm \int \dy f(y) \d_{\mm+\mu}\delta(y)\, \nl
&=& -\summp \frac1{\mm!}\, x^\mm \int \dy \d_{\mm+\mu} f(y)\, \delta(y) 
\nl
&=& -\summp \frac1{\mm!}\, \d_{\mm+\mu} f(0)\, x^\mm
= -\dmu f(x) |_p
\eens

\section{Some infinite sums}
\bes
&i.& A_{d,p} \equiv \summp 1 
= \binom{d+p}{d}, 
\label{Adp}\\
&ii.& B_{d,p} \equiv \summp m_\mu 
= \binom{d+p}{d+1}, 
\label{Bdp}\\
&iii.& C_{d,p} \equiv \summp m_\mu^2 
= \binom{d+p}{d+2} + \binom{d+p+1}{d+2}, 
\label{Cdp}\\
&iv.& D_{d,p} \equiv \summp m_\mu m_\nu 
= \binom{d+p}{d+2}, \qquad \hbox{if $\mu \neq \nu$}, 
\label{Ddp}\\
&v.& E_{d,p} \equiv \summp m_\mu (m_\nu + 1)
= \binom{d+p+1}{d+2}, \qquad \hbox{if $\mu \neq \nu$}, 
\label{Edp}\\
\eens

\noindent {\em Proof:} The relations are proven by induction, repeatedly using 
the recursion formula (\cite{AS65}, 2.1.1 II):
\bes
\binom{n+1}{m} &=& \binom{n}{m} + \binom{n}{m-1} \nl
&=& \binom{n}{m} + \binom{n-1}{m-1} + ... + \binom{n-m}{0} \\
&=& \binom{n}{n-m} + \binom{n-1}{n-m} + ... + \binom{n-m}{n-m}.
\eens

\smallskip
\noindent $i$. 
Clearly, $A_{1,p} = \sum_{m=0}^p 1 = p+1$.
For the recursive step,
\bes
A_{d,p} &=& \sum_{i=0}^p \sum_{|\mm|\leq p-i} 1
= \sum_{i=0}^p A_{d-1,p-i} \nl
&=& \sum_{i=0}^p \binom{d-1+p-i}{d-1}
= \sum_{i=0}^p \binom{n-i}{n-p} \\
&=& \binom{n+1}{p}
= \binom{d+p}{p}
= \binom{d+p}{d},
\eens
where $\mm$ is a multi-index with $d-1$ components,
$n = d+p-1$, and we used the recursion formula.

\smallskip
\noindent $ii$. 
For $d=1$, $B_{1,p} = \sum_{m=0}^p m = \binom{p+1}{2}$.
Let $i = m_{d-1}$ be the last component, and let $\mm$ denote a multi-index
with $d-1$ components.
\bes
B_{d,p} &=& \sum_{i=0}^p \sum_{|\mm|\leq p-i} i
= \sum_{i=0}^p i A_{d-1,p-i} \nl
&=& \sum_{j=0}^{p-1} (j+1) A_{d-1,p-j-1} \\
&=& \sum_{j=0}^{p-1} j A_{d-1,p-j-1} + \sum_{j=0}^{p-1} A_{d-1,p-j-1}.
\eens
Thus
\be
B_{d,p} = B_{d,p-1} + A_{d,p-1} = B_{d,p-1} + \binom{d+p-1}{d}.
\ee
Using the recursion formula we verify that 
$B_{d,p} = \binom{d+p}{d+1}$.

\smallskip
\noindent $iii$. 
When $d=1$, 
\be
C_{1,p} = \sum_{m=0}^p m^2 = \binom{p+1}3 + \binom{p+2}3 = \frac16(p+3p^2+2p^3), 
\ee
which is proven with a cubic ansatz for $C_{1,p}$ and 
identification of components. The recursive step is
\bes
C_{d,p} &=& \sum_{i=0}^p \sum_{|\mm|\leq p-i} i^2
= \sum_{i=0}^p i^2 A_{d-1,p-i} \nl
&=& \sum_{j=0}^{p-1} (j^2 + 2j + 1) A_{d-1,p-j-1} 
\\
&=& C_{d,p-1} + 2B_{d,p-1} + A_{d,p-1}
= C_{d,p-1} + 2B_{d,p} - A_{d,p-1} \nl
&=& C_{d,p-1} + 2\binom{d+p}{d+1} - \binom{d+p-1}{d} \nl
&=& C_{d,p-1} + \binom{d+p}{d+1} + \binom{d+p-1}{d+1}.
\eens
We use the recursion formula to verify that the given expression 
for $C_{d,p}$ indeed satisfies this relation.

\smallskip
\noindent $iv$. 
Let $i = m_{d-1}$, $j = m_{d-2}$, and let $\mm$ have $d-2$ components.
\bes
D_{d,p} &=& \sum_{i=0}^p \sum_{j=0}^{p-i} \sum_{|\mm|\leq p-i-j} ij 
= \sum_{i=0}^p \sum_{j=0}^{p-i} ij A_{d-2,p-i-j} \nl
&=& \sum_{i=0}^p i B_{d-1,p-i}
= \sum_{k=0}^{p-1} (k+1) B_{d-1,p-k-1} 
\nle
&=& D_{d,p-1} + \sum_{k=0}^{p-1} \binom{d+p-2-k}{d} \nl
&=& D_{d,p-1} + \binom{d+p-1}{p-2} 
= D_{d,p-1} + \binom{d+p-1}{d+1}.
\eens
In the second last step, we made the substitution
$n = d-2+p$, $n-m = d$, $m = p-2$, and used the recursion formula to
calculate the sum of binomial coefficients. We verify that
$D_{d,p} = \binom{d+p}{d+2}$ satisfies this relation.

\smallskip
\noindent $v$. 
\bes
E_{d,p} &=& D_{d,p} + B_{d,p} \nl
&=& \binom{d+p}{d+2} + \binom{d+p}{d+1}
= \binom{d+p+1}{d+2}.
\ees

\section{Evaluation of products of $p$-jet delta functions}
\label{app:delta_product}

Unlike the ordinary delta function $\delta(x)$, the $p$-jet delta
functions can be multiplied with each other in a meaningful way. 
In this appendix we evaluate three expressions that are needed in the text.

Let $f(x)$ and $g(x)$ be some smearing functions, and denote by
$f_0(x) = f(x) - f(0)$ and $g_0(x) = g(x) - g(0)$ the corresponding
shifted functions. Clearly,
$f_0(0) = g_0(0) = 0$. Since the functions are just shifted by a constant,
the suffix $0$ is not necessary in derivatives:
$\dmu f_0(x) = \dmu f(x)$.

The following sums are needed in the text:
\bes
&i.& \iint \dx\dy f(x)g(y)\, \delta_p(x,y)\, \delta_p(y,x) \nl
&&\, =\, \binom{d+p}{d} f(0)g(0), 
\label{sum1}\\
&ii.& \iint \dx\dy f_0(x)g(y)\, \d^x_\mu \delta_p(x,y)\, \delta_p(y,x) \nl
&&\, =\, \binom{d+p}{d+1} \dmu f(0)g(0), 
\label{sum2}\\
&iii.& \iint \dx\dy f_0(x)g_0(y)\, \d^x_\mu \delta_p(x,y)\, \d^y_\nu \delta_p(y,x) \nl
&&\, =\, \binom{d+p+1}{d+2} \dnu f(0) \dmu g(0)
+ \binom{d+p}{d+2} \dmu f(0) \dnu g(0).
\label{sum3}
\ees
{\em Proof:}

\smallskip
\noindent $i$.
By definition of the $p$-jet delta function, the LHS reads
\bes
&& \iint \dx\dy f(x)g(y)\, \sum_m \mfrac x^\mm \d_\mm\delta(y) 
 \, \sum_n \nfrac y^\nn \d_\nn\delta(x) 
\nl
&=& \sum_{\mm,\nn} \frac1{\mm!\nn!} \iint \dx\dy
 \d_\nn(x^\mm f(x))\, \d_\mm(y^\nn g(y))\, \delta(x) \delta(y),
\label{t1}
\ees
where we integrated by parts $\mm$ times in $x$ and $\nn$ times in $y$. Now,
\be
\int \dx \d_\nn(x^\mm f(x))\, \delta(x) = 0,
\ee
unless $\nn \geq \mm$ (which means that $n_\mu > m_\mu$ for all $\mu$), because 
otherwise the integrand would be propotional to $x^\mu$, and the delta function
would kill it. At the same time, we must also have $\mm \geq \nn$, because 
otherwise the integration over $y$ would vanish. Hence $\mm=\nn$ and (\ref{t1}) is equal to
\bes
&& \sum_\mm \frac1{(\mm!)^2} \iint \dx\dy
 \, \d_\mm(x^\mm f(x))\, \d_\mm(y^\mm g(y))\, \delta(x) \delta(y) \nl
&=& \sum_\mm \frac1{(\mm!)^2} \iint \dx\dy (\mm! f(x))\, (\mm! g(y))\, \delta(x) \delta(y) \nl
&=& \sum_\mm 1 \cdot f(0) g(0) \\
&=& A_{d,p}\, f(0) g(0),
\eens
where $A_{d,p}$ is defined in (\ref{Adp}).

\smallskip
\noindent $ii$.
The LHS becomes
\bes
&&\iint\dx\dy f_0(x)g(y) \sum_\mm \mfrac m_\mu x^{\mm-\mu} \d_\mm\delta(y)
 \sum_\nn \nfrac y^\nn \d_\nn \delta(x) 
\nle
&=& \sum_{\mm,\nn} \frac1{(\mm-\mu)!\nn!} \iint \dx\dy
 \d_\nn(x^{\mm-\mu} f_0(x))\, \d_\mm(y^\nn g(y))\, \delta(x) \delta(y),
\eens 
after repeated integration by parts.
In order to kill off all factors of $x^\mu$, we must have $\nn \geq \mm-\mu$. 
There are two ways this can be achieved:
\bes
&a)& \mm = \nn, \nl
&b)& \nn = \mm-\mu, \qquad \mm = \nn+\mu.
\eens
However, in the second case the expression is proportional to
\be
\int \dx \d_\nn(x^\nn f_0(x))\, \delta(x) = \nn! f_0(0) = 0,
\ee
because the smearing function $f_0(x)$ is shifted to make $f_0(0) = 0$.
Hence only the first case contributes to the result, which becomes
\bes
&& \sum_\mm \frac1{(\mm-\mu)!\mm!} \iint \dx\dy
 \, m_\mu (\mm-\mu)! \dmu f_0(x)\, \mm!g(y)\, \delta(x) \delta(y) \nl
&=& \sum_\mm m_\mu\cdot \dmu f_0(0) g(0) \\
&=& B_{d,p}\, \dmu f(0) g(0),
\eens
where we used (\ref{Bdp}) and $\dmu f_0(0) = \dmu f(0)$ in the last step.

\smallskip
\noindent $iii$.
The LHS becomes
\bes
&&\iint\dx\dy f_0(x)g_0(y) \sum_\mm \mfrac m_\mu x^{\mm-\mu} \d_\mm\delta(y) \times \nl
 &&\qquad\times\, \sum_\nn \nfrac n_\nu y^{\nn-\nu} \d_\nn \delta(x) 
\nlb{t3}
&=& \sum_{\mm,\nn} \frac1{(\mm-\mu)!(\nn!-\nu)} \times \nl
 &&\qquad\times\, \iint \dx\dy
 \d_\nn(x^{\mm-\mu} f_0(x))\, \d_\mm(y^{\nn-\nu} g_0(y))\, \delta(x) \delta(y).
\eens 
There are now five cases that may survive:
\bes
&a)& \mm = \nn, \nl
&b)& \nn = \mm-\mu, \qquad \mm = \nn+\mu, \nl
&c)& \mm = \nn-\nu, \qquad \nn = \mm+\nu, \nl
&d)& \mu \neq \nu \mbox{ and } \nn = \mm-\mu+\nu, \mm = \nn+\mu-\nu, \nl
&e)& \mu = \nu \mbox{ and } \nn = \mm-2\mu, \mm = \nn+2\mu.
\eens
In case a), (\ref{t3}) equals
\bes
&& \sum_\mm \frac1{(\mm-\mu)!(\mm-\nu)!}\, 
 m_\mu (\mm-\mu)!\, \dmu f_0(0)\, m_\nu (\mm-\nu)!\, \dnu g_0(0) \nl
&=& \sum_\mm m_\mu m_\nu\cdot \dmu f_0(0) \dnu g_0(0) \\
&=& \begin{cases}
 C_{d,p}\, \dmu f(0) \dnu g(0), & \mbox{if } \mu = \nu, \\
 D_{d,p}\, \dmu f(0) \dnu g(0), & \mbox{if } \mu \neq \nu,
\end{cases}
\eens
where the sum was evaluated in (\ref{Cdp}) and (\ref{Ddp}).

\noindent
Cases b) and c) both vanish, because the expressions are proportional to
\bes
\int \dx \d_\nn(x^\nn f_0(x)) \delta(x) = \nn! f_0(0) = 0, \nle
\int \dy \d_\mm(y^\mm g_0(y)) \delta(y) = \mm! g_0(0) = 0,
\eens
respectively.

Case d) is already covered by case a) if $\mu=\nu$. If $\mu\neq \nu$,
equation (\ref{t3}) reads
\bes
&& \sum_\mm \frac1{(\mm-\mu)!(\nn-\nu)!} \iint \dx\dy
 \d_\nn(x^{\nn-\nu} f_0(x))\, \d_\mm(y^{\mm-\mu} g_0(y))\, \delta(x) \delta(y) \nl
&=& \sum_\mm n_\nu \dnu f_0(0)\, m_\mu \dmu g_0(0) \\
&=& \sum_\mm m_\mu (\mm-\mu+\nu)_\nu\, \dnu f(0) \dmu g(0).
\eens
We now note that $\nu_\nu = 1$ and $\mu_\nu = 0$ because $\mu \neq \nu$.
Using (\ref{Edp}) we arrive at 
\be
\sum_\mm m_\mu (m_\nu+1)\, \dnu f(0) \dmu g(0)
= E_{d,p}\, \dnu f(0) \dmu g(0).
\ee

\noindent
Finally we have case e), but this vanishes because it is proportional to
\be
\int \dx \d_\nn(x^{\nn+2\mu} f_0(x)) \delta(x) = 0,
\ee
and there are not enough derivatives to kill all powers of $x$.

Summing up the non-zero contributions from cases a) and d), the result for
the integral in (\ref{t3}) is
\be
C_{d,p}\, \dmu f(0) \dnu g(0)
\label{r3a}
\ee
if $\mu = \nu$, and
\be
E_{d,p}\, \dnu f(0) \dmu g(0) + D_{d,p}\, \dmu f(0) \dnu g(0)
\label{r3b}
\ee
if $\mu \neq \nu$. However, since $C_{d,p} = E_{d,p} + D_{d,p}$, the
latter expression is equal to (\ref{r3a}) when $\mu=\nu$, so the covariant result
(\ref{r3b}) holds irrespective of whether $\mu$ equals $\nu$ or not.

\section{ Evaluation of the $Vir(d)$ OPE (\ref{LLzw})}
\label{app:LL}

In this appendix we evaluate the OPE between the three partial
$Vir(d)$ generators listed in (\ref{Lixi}). Since none of the
generators involves any derivatives of $z$, the Wick contractions
must be of the form
\be
L^i_\xi(z) L^j_\eta(w) \sim \frac{R^{ij}_\xeta(w)}{(z-w)}
+ \frac{Z^{ij}_\xeta(w)}{(z-w)^2}.
\ee
To reduce writing, we suppress arguments where so can be done without
obscuring the meaning.

First consider the regular terms.
\bes
R^{00}_\xeta &=& \no{\xmu(q) p_\mu} \no{\ynu(q) p_\nu} \nl
&\sim& \ynu(q)\dnu\xmu(q) p_\mu - \xmu(q)\dmu\ynu(q) p_\mu \nl
&=& L^0_{[\xi,\eta]}
\label{R00}\\
R^{01}_\xeta &=& -\no{\xmu(q) p_\mu} \int(\ynu(x+q) - \ynu(q)) \no{\pi\dnu\phi} \nl
&\sim& \xmu(q) \int(\dmu\ynu(x+q) - \dmu\ynu(q)) \no{\pi\dnu\phi}
\label{R01}\\
R^{10}_\xeta &=& -\int (\xmu(x+q) - \xmu(q))\no{\pi\dmu\phi} \no{\ynu(q) p_\nu} \nl
&\sim& -\ynu(q) \int (\dnu\xmu(x+q) - \dnu\xmu(q))\no{\pi\dmu\phi}
\label{R10}
\ees
To evaluate $R^{11}_\xeta$, define
\be
\xmu_0(x) = \xmu(x+q) - \xmu(q)
\ee
We then have
\bes
R^{11}_\xeta &=& \iint \xmu_0(x)\ynu_0(y) \no{\pi(x)\dmu\phi(x)}\no{\pi(y)\dnu\phi(y)} \nl
&\sim& \iint \xmu_0(x)\ynu_0(y) \Big(
 \d^x_\mu\delta_p(x,y) \no{\pi(x)\dnu\phi(y)} \nl
 &&-\ \d^y_\nu\delta_p(y,x) \no{\pi(y)\dnu\phi(x)} 
 \Big) 
\label{R11} \\
&=& \int \xmu_0 \dmu\ynu \no{\pi\dnu\phi} + \int \xmu_0 \ynu_0 \no{\pi\dmu\dnu\phi} \nl
 &&-\ \int \ynu_0\dnu\xmu \no{\pi\dmu\phi} - \int \xmu_0 \ynu_0 \no{\pi\dnu\dmu\phi} \nl
&=& \int (\xmu(x+q)-\xmu(q))\dmu\ynu(x+q) \no{\pi\dnu\phi} \nl
 &&-\ \int (\ynu(x+q)-\ynu(q))\dnu\xmu(x+q) \no{\pi\dmu\phi}.
\eens
Partial integration in the second step was performed using (\ref{partint_x}).
The sum of (\ref{R01}), (\ref{R10}) and (\ref{R11}) is
\bes
R^{01}_\xeta + R^{10}_\xeta + R^{11}_\xeta
&=& \int (\xmu\dmu\ynu)_0 \no{\pi\dnu\phi} 
 - \int (\ynu\dnu\xmu)_0 \no{\pi\dmu\phi} \nl
&=& L^1_{[\xi,\eta]}.
\label{R1}
\ees
The next two terms are
\bes
R^{02}_\xeta &=& -\no{\xmu(q) p_\mu} \int \drho\ynu(x+q) \no{\pi T^\rho_\nu \phi} \nl
&\sim& \xmu(q) \int \dmu\drho\ynu(x+q)\no{\pi T^\rho_\nu \phi}
\label{R02}\\
R^{12}_\xeta &=& \iint \xmu_0(x)\drho\ynu(y) \no{\pi(x)\dmu\phi(x)} \no{\pi(y) T^\rho_\nu \phi(y)} \nl
&\sim& \iint \xmu_0(x)\drho\ynu(y) \Big(
 \d^x_\mu\delta_p(x,y)\no{\pi(x) T^\rho_\nu \phi(y)} \nl
 && -\ \delta_p(y,x)\no{\pi(y) T^\rho_\nu \dmu\phi(x)} 
 \Big) 
\label{R12} \\
&=& \int \Big( \xmu_0\dmu\drho\ynu\no{\pi T^\rho_\nu\phi}
 +\ \xmu_0\drho\ynu(y)\no{\pi T^\rho_\nu\dmu\phi} \nl
 &&-\ \xmu_0\drho\ynu(y)\no{\pi T^\rho_\nu\dmu\phi} \Big) \nl
&=& \int (\xmu(x+q)-\xmu(q))\dmu\drho\ynu\no{\pi T^\rho_\nu\phi}
\eens
Summing the last two contribution, we find
\be
R^{02}_\xeta + R^{12}_\xeta = \int \xmu(x+q)\dmu\drho\ynu(x+q) \no{\pi T^\rho_\nu\phi}
\ee
and analogously
\be
R^{20}_\xeta + R^{21}_\xeta = -\int \ynu(x+q)\dnu\drho\xmu(x+q) \no{\pi T^\rho_\mu\phi}
\ee
Finally,
\bes
R^{22}_\xeta &=& \iint \drho\xmu\dsi\ynu \no{\pi T^\rho_\mu \phi}\no{\pi T^\si_\nu \phi} \nl
&\sim& \int \drho\xmu\dsi\ynu \no{\pi [T^\rho_\mu, T^\si_\nu] \phi} 
\label{R22} \\
&=& \int \drho\xmu\dmu\ynu \no{\pi T^\rho_\nu\phi}
 - \int \drho\ynu\dnu\xmu \no{\pi T^\rho_\mu\phi}
\eens
Summing the last few constribution, we find that the sum
$R^{02}_\xeta + R^{12}_\xeta +R^{20}_\xeta + R^{21}_\xeta + R^{22}_\xeta$ equals
\bes
&&\int (\drho\xmu\dmu\ynu + \xmu\dmu\drho\ynu) \no{\pi T^\rho_\nu\phi} \nl
&&-\ \int (\drho\ynu\dnu\xmu + \ynu\dnu\drho\xmu) \no{\pi T^\rho_\mu\phi}
\label{R2} \\
&=& \int \drho [\xi,\eta]^\mu \no{\pi T^\rho_\mu\phi} 
\ = L^2_{[\xi,\eta]}
\eens
Using (\ref{R00}), (\ref{R1}) and (\ref{R2}), we finally have
\be
\sum_{i=0}^2 \sum_{j=0}^2 R^{ij}_\xeta 
= L^0_{[\xi,\eta]} + L^1_{[\xi,\eta]} + L^2_{[\xi,\eta]}
= L_{[\xi,\eta]}.
\ee

We now turn to the double Wick contractions.
\bes
Z^{00}_\xeta &=&
 \bcontraction [2ex] {:\xmu(} {q} {)p_\mu::\ynu(q)} {p} 
 \bcontraction{:\xmu(q)} {p} {_\mu::\ynu(} {q}
  \no{\xmu(q)p_\mu} \no{\ynu(q)p_\nu} 
= -\dnu\xmu(q) \dmu\ynu(q)
\label{Z00}
\ees
The other terms involving the observer's momentum vanish,
\be
Z^{01}_\xeta(z,w) = Z^{10}_\xeta(z,w) 
= Z^{02}_\xeta(z,w) = Z^{20}_\xeta(z,w) = 0,
\ee
because in these terms only a single Wick contraction is possible.
\bes
Z^{11}_\xeta &=& \iint \xmu_0(x)\ynu_0(y)
 \bcontraction[2ex] {:} {\pi} {(x)\dmu\phi(x)::\pi(y)\dnu} {\phi}
 \bcontraction {:\pi(x)\dmu} {\phi} {(x)::} {\pi}
 \no{\pi(x)\dmu\phi(x)} \no{\pi(y)\dnu\phi(y)} \nl
&=& \mp \tr\,I\, \iint \xmu_0(x)\ynu_0(y)\, \d^x_\mu\delta_p(x,y) \d^y_\nu\delta_p(y,x)
\label{Z11}\\
&=& \mp \Delta_\varrho \Big( E_{d,p} \dnu\xmu(q)\dmu\ynu(q)
 +\ D_{d,p}\dmu\xmu(q)\dnu\ynu(q) \Big),
\eens
where we used (\ref{sum3}) and the fact that there are
$\Delta_\varrho = \tr\, I$ different fields that contribute to
the sum.
\bes
Z^{12}_\xeta &=& \iint \xmu_0(x)\drho\ynu(y)
 \bcontraction[2ex] {:} {\pi} {(x)\dmu\phi(x): :\pi(y)T^\rho_\nu} {\phi} 
 \bcontraction {:\pi(x)\dmu} {\phi} {(x)::} {\pi} 
 \no{\pi(x)\dmu\phi(x)} \no{\pi(y)T^\rho_\nu\phi(y)} \nl
&=& \mp\tr\,T^\rho_\nu\,\iint \xmu_0(x)\drho\ynu(y)\,  
 \d^x_\mu\delta_p(x,y)\delta_p(y,x)
\label{Z12}\\
&=& \mp B_{d,p} k_0(\varrho)\, \dmu\xmu(q)\, \dnu\ynu(q)
\eens
Here we used (\ref{sum2}) to evaluate the sum and that the trace of
the $\gl(d)$ generators $\tr\,T^\mu_\nu = k_0(\varrho) \delta^\mu_\nu$.
By symmetry we immediately obtain $Z^{21}_\xeta = Z^{12}_\xeta$.
\bes
Z^{22}_\xeta &=& \iint \drho\xmu(x)\dsi\ynu(y)
 \bcontraction[2ex] {:} {\pi} {(x)T^\rho_\mu\phi(x): :\pi(y)T^\si_\nu} {\phi} 
 \bcontraction {:\pi(x)T^\rho_\mu} {\phi} {(x): :} {\pi} 
 \no{\pi(x)T^\rho_\mu\phi(x)} \no{\pi(y)T^\si_\nu\phi(y)} \nl
&=& \mp \tr\,T^\rho_\mu T^\si_\nu \iint \drho\xmu(x)\dsi\ynu(y)\, 
 \delta_p(x,y)\delta_p(y,x)
\label{Z22} \\
&=& \mp A_{d,p}\, (k_1(\varrho) \delta^\rho_\nu\delta^\si_\mu 
 + k_2(\varrho) \delta^\rho_\mu\delta^\si_\nu)\, \drho\xmu(q) \dsi\ynu(q),
\eens
where we used (\ref{sum1}) and (\ref{glparam}). 

Summing the nonzero contributions, 
\be
Z_\xeta = Z^{00}_\xeta + Z^{11}_\xeta + 2Z^{12}_\xeta + Z^{22}_\xeta,
\ee
we see that the extension is of the form (\ref{Zxieta}) and the abelian
charges are given by (\ref{c1c2}).

\section{Proof of the results in subsection \ref{ssec:VirAff}}
\label{app:VirAff}

The OPE will be of the form
\be
L_\xi(z) J_X(w) \sim \frac{H_\xX(w)}{z-w} + \frac{W_\xX(z,w)}{(z-w)^2}.
\ee
First we calculate the regular term:
\bes
H_\xX(w) &=& \xmu(q) \int \dmu X^a \no{\pi M^a\phi}
+ \int X^a \no{(-\xmu_0\dmu\pi + \dnu\xmu\pi T^\nu_\mu) M^a\phi} \nl
&&+\ \int X^a \no{\pi M^a(-\xmu_0\dmu\phi - \dnu\xmu T^\nu_\mu)\phi} \\
&=& \xmu(q)\int \Big( \dmu X^a\no{\pi M^a\phi} 
+ X^a\dmu(\no{\pi M^a\phi}) \Big) \nl
&&-\ \int X^a\xmu \dmu (\no{\pi M^a\phi}).
\eens
The first term is a total derivative, and the second becomes after an
integration by parts:
\be
H_\xX(w) = \int \dmu(\xmu(x+q)X^a(x+q))\no{\pi M^a\phi}.
\ee
We recognize the transformation law for a density of weight one.

The extension only involves contributions from $L^1_\xi(z)$ and $L^2_\xi(z)$,
because there is no double contraction between $L^0_\xi(z)$ and $J_X(w)$.
\bes
W_\xX &=& \iint \Big( \xmu_0(x) X^a(y)
 \bcontraction[2ex] {:} {\pi} {(x)\dmu\phi(x): :\pi(y)M^a} {\phi} 
 \bcontraction {:\pi(x)\dmu} {\phi} {(x): :} {\pi}
 \no{\pi(x)\dmu\phi(x)} \no{\pi(y) M^a\phi(y)} \nl
&&+\ \dnu\xmu(x) X^a(y)
 \bcontraction[2ex] {:} {\pi} {(x)T^\nu_\mu\phi(x): :\pi(y)M^a} {\phi} 
 \bcontraction {:\pi(x)T^\nu_\mu} {\phi} {(x): :} {\pi}
 \no{\pi(x)T^\nu_\mu\phi(x)} \no{\pi(y) M^a\phi(y)} \nl
&=& \mp \tr\,M^a \iint \Big(
 \xmu_0 X^a\, \d^x_\mu\delta_p(x,y)\delta_p(y,x) \,\tr\,I \\
 &&+\ \dnu\xmu X^a\, \delta_p(x,y)\delta_p(y,x) \,\tr\,T^\mu_\nu \Big) \nl
&=& \mp z_M\delta^a \Big( B_{d,p} \Delta_\varrho\, \dmu\xmu(q) X^a(q)
 + A_{d,p} k_0(\varrho) \delta^\nu_\mu\, \dnu\xmu(q) X^a(q) 
 \Big) \nl
&=& \mp z_M \Big( B_{d,p} \Delta_\varrho + A_{d,p} k_0(\varrho) \Big)
  \dmu\xmu(q) \delta^a X^a(q),
\eens
where we used (\ref{sum1}) and (\ref{sum2}) to evaluate the products of
delta functions. The form of the OPE and the expression for the 
abelian charge $c_7$ (\ref{WxX}) -- (\ref{c7}) are thus confirmed.

\section{ Proof of the equations in section \ref{sec:repar} }
\label{app:repar}

Since the reparametrization generators depend on $z$-derivatives, it is
useful to list the OPEs between $z$-derivatives of the fundamental fields:
\bes
\phi(x,z)\pi(y,w) \sim \frac{1}{z-w}\delta_p(x,y)
&&
\pi(x,z)\phi(y,w) \sim \frac{\mp1}{z-w}\delta_p(y,x)
\nl
\dot\phi(x,z)\pi(y,w) \sim \frac{-1}{(z-w)^2}\delta_p(x,y)
&&
\pi(x,z)\dot\phi(y,w) \sim \frac{\mp1}{(z-w)^2}\delta_p(y,x)
\nl
\phi(x,z)\dot\pi(y,w) \sim \frac{1}{(z-w)^2}\delta_p(x,y)
&&
\dot\pi(x,z)\phi(y,w) \sim \frac{\pm1}{(z-w)^2}\delta_p(y,x)
\nl
\dot\phi(x,z)\dot\pi(y,w) \sim \frac{-2}{(z-w)^3}\delta_p(x,y)
&&
\dot\pi(x,z)\dot\phi(y,w) \sim \frac{\pm2}{(z-w)^3}\delta_p(y,x)
\eens

In general, the $T(z)T(w)$ extension is obtained by double contractions of
four terms. For brevity we only consider the case $\lambda=0$ where the extension 
only consists of a single term:
\bes
T(z)T(w) &\sim& \iint
 \bcontraction[2ex] {:} {\pi} {(x,z)\phi(x,z)::\pi(y,w)} {\phi}
 \bcontraction {:\pi(x,z)} {\phi} {(x,z): :} {\pi}
 \no{\pi(x,z)\dot\phi(x,z)}\no{\pi(y,w)\dot\phi(y,w)} \nl
&=& \tr\,I \iint \frac{-1}{(z-w)^2}\delta_p(x,y)\, \frac{\mp1}{(z-w)^2} \delta_p(y,x) \\
&=& \frac{\pm 1}{(z-w)^4} \Delta_\varrho \Delta_M A_{d,p}.
\eens
The results for non-zero $\lambda$ is obtained analogously. The central charge
$c_4$ simply equals the usual Virasoro central charge (\ref{clambda}) times the
number of conformal fields $\binom{d+p}{d} \Delta_\varrho \Delta_M$.

The extension in the $T(z)J_X(w)$ OPE is
\bes
&& \int ((\lambda-1)\no{\pi\dot\phi} + \lambda\no{\dot\pi\phi} )
  \int \no{\pi X\phi} \nl
&\sim& \iint \tr\,X \frac{\pm(2\lambda-1)}{(z-w)^3}\,
  \tr\,I\, \delta_p(x,y) \delta_p(y,x) 
\nle
&=& \pm\tr\,X(q)\, \frac{2\lambda-1}{(z-w)^3}\,
 \Delta_\varrho A_{d,p} \nl
&=& \pm z_M \delta^a X^a(q)\, \frac{2\lambda-1}{(z-w)^3}\, \Delta_\varrho A_{d,p},
\eens
which is (\ref{TJ-ope}).

The extension of the $T(z)L_\xi(w)$ OPE is
\bes
&& \int ((\lambda-1)\no{\pi\dot\phi} + \lambda\no{\dot\pi\phi} )
\int (\xmu_0\no{\pi\dmu\phi} + \dnu\xmu \no{\pi T^\nu_\mu\phi} \nl
&\sim& \frac{\pm(2\lambda-1)}{(z-w)^3} \iint \Big(
 \xmu_0\, \tr\,I\, \delta_p(x,y) \d^y_\mu \delta_p(y,x) 
 + \dnu\xmu\, \tr\,T^\nu_\mu\, \delta_p(x,y)\delta_p(y,x) 
 \Big)\nl
&=& \pm\frac{2\lambda-1}{(z-w)^3} \Big( \dmu\xmu(q) \Delta_\varrho B_{d,p} 
 + \dmu\xmu(q) k_0(\varrho) A_{d,p} \Big)\Delta_M,
\ees
which is (\ref{TL-ope}).

\end{document}